\documentclass{article}
\usepackage{sw20lart}

\input{tcilatex}
\begin{document}

\title{Change, Time and Information Geometry\thanks{%
Presented at MaxEnt 2000, the 20th International Workshop on Bayesian
Inference and Maximum Entropy Methods (July 8-13, 2000, Gif-sur-Yvette,
France).}}
\author{Ariel Caticha \\
{\small Department of Physics, University at Albany-SUNY, }\\
{\small Albany, NY 12222, USA.\thanks{%
E-mail: ariel@cnsvax.albany.edu or Ariel.Caticha@albany.edu}}}
\date{}
\maketitle

\begin{abstract}
Dynamics, the study of change, is normally the subject of mechanics. Whether
the chosen mechanics is ``fundamental'' and deterministic or
``phenomenological'' and stochastic, all changes are described relative to
an external time. Here we show that once we define what we are talking
about, namely, the system, its states and a criterion to distinguish among
them, there is a single, unique, and natural dynamical law for irreversible
processes that is compatible with the principle of maximum entropy. In this
alternative dynamics changes are described relative to an internal,
``intrinsic'' time which is a derived, statistical concept defined and
measured by change itself. Time is quantified change.
\end{abstract}

\section{Introduction}

The notion that the concepts of time, change and motion are intimately
connected goes back to antiquity. According to Aristotle, ``time numbers
change with respect to before and after.'' One aspect of this connection is
the order of a sequence of changes, their temporal order. Another aspect is
the use of selected motions or changes to measure the length of time
intervals, their duration. We begin by considering the notion of change.

In order to establish that a system has changed one must be able to
distinguish between the system being in one state and its being in another
state. This requires, to begin with, a clear idea of what is meant by a
state. As long as one is interested in the study of phenomena that can be
deliberately reproduced by controlling a few macroscopic variables it is
reasonable to expect that the values -- or rather, the expected values -- of
these few variables are all that is needed for the purposes of prediction.
This limited information defines what we mean by the state or, equivalently,
the macrostate of the system.

Next, to measure the extent to which states can be distinguished, we assign
a probability distribution to each state. The requirement that the
assignment procedure itself do not introduce any information beyond that
which defines the state demands we use the method of maximum entropy (ME) 
\cite{Jaynes57}\cite{Skilling88}. In this way the problem of distinguishing
between states is transformed into another problem, that of distinguishing
between the corresponding distributions. The solution to the latter problem
is well known. There is a uniquely natural way to quantify the extent to
which one distribution can be distinguished from another: it is given by the
distance between them as measured by the Fisher-Rao information metric \cite
{Fisher25}-\cite{Rodriguez89}.

If we think of each state as a point in a manifold, the net outcome of these
considerations (Sect. 2) is that the method of ME has transformed the
manifold of states into a metric space. Distinguishability and therefore
change is measured by distance.

There is not yet any implication that change will happen \emph{from} one
state \emph{to} another; to this we turn next. Temporal order, as well as
the notion of time itself, are the subject of dynamics.

Typically, having decided on the kinematics appropriate to a certain motion,
one defines the dynamics by additional postulates about the equations of
motion, perhaps in the form of a variational principle. The dynamics is
postulated. The dynamical law that we adopt here (Sect. 3) is a variational
principle too, but there is something very peculiar about it, there is no
need to postulate it. The principle is the same we had already introduced
when discussing the space of states, namely, when selecting a distribution
subject to certain constraints, the preferred distribution is that of
maximum entropy. It is just the same old ME principle applied in a somewhat
different way. (The nature of the constraints is different. For a brief
account of the ME method in a form tailored to suit the needs of this paper
see Ref.\cite{Caticha00}.)

We have no freedom in choosing the dynamical law; it follows from the single
piece of new information available: recognizing that changes happen. Nothing
else. Suppose the system is in a certain state and a small change happens;
the system moves a distance $d\ell $. We cannot with certainty predict in
which direction motion occurs but, according to the principle of ME, unless
there is some positive evidence to the contrary, of all the states on the
surface of the sphere of radius $d\ell $ there is one to be preferred above
all others: it is the state of maximum entropy.

As so often in the past, it seems that once more the method of ME has
allowed us to get something out of nothing; yet another free lunch. But the
dynamics proposed here is different in one important respect. (We refrain
from saying ``defficient'' rather than ``different'' because in the end it
may turn out to be an advantage.) In the conventional Hamiltonian or
Lagrangian mechanics the equations of motion describe changes relative to an
external time. Here changes are described relative to an internal,
``intrinsic'' time which is a derived, statistical concept defined and
measured by the change $d\ell $ itself. Intrinsic time is quantified change.
The system provides its own clock. Perhaps this is a necessary feature of
any fundamental form of mechanics that generates its own notion of time,
that \emph{explains} time.

The introduction of a metric in the space of states is not new; this has
been done by many authors in statistical inference, where the subject is
known as Information Geometry \cite{Amari85}\cite{Rodriguez90}, and in
physics, to study both equilibrium \cite{Weinhold75}\cite{Ingarden76} and
nonequilibrium thermodynamics \cite{Balian86}\cite{Streater95}. What is
different here is the recognition that this is all one needs to define a
dynamics.

An interesting consequence of these ideas is that reciprocity relations of
the Onsager type \cite{Onsager31} valid near and far from equilibrium are
obtained (Sect. 4) without any hypothesis about microscopic reversibility;
in fact, no mention is made of any microscopic dynamics. By analyzing
specific models other authors \cite{Gabrielli96} have reached similar
conclusions: reciprocal relations are possible even if the underlying
microscopic dynamics is not reversible.

It is, of course, possible to incorporate more information, that is,
additional constraints into the dynamics. In Sect. 5 we consider a simple
illustrative example, the intrinsic dynamics of two coupled systems as they
evolve towards equilibrium along a trajectory constrained by conservation
laws.

Our subject can be approached from another direction. The Greeks did not
draw a sharp distinction between change in general and the more special kind
of change we call motion; the falling of an apple was not viewed as being in
any sense more fundamental than the ripening of an apple. The modern view
does draw such distinctions; deterministic motion in space and time is
considered basic while other kinds of change -- notably irreversible
processes in macroscopic systems -- are not. They must be understood in
terms of the deterministic motion of microscopic constituents. Of course,
this view is not wrong, but for some purposes it may be misguided,
inconvenient.

All theories describing irreversible processes have, in the past, invariably
turned out to be rather formidable (see e.g., \cite{Grabert82}-\cite{Luzzi90}%
). One reason is that the phenomena to be described are themselves quite
complicated. But there is another reason, which is that these theories are
attempting to achieve two conflicting goals. \smallskip One goal is to reach
an understanding in terms of the microscopic Hamiltonian laws of motion and
requires keeping track of microscopic details. The other goal is to achieve
a description in terms of the few variables that matter, those that codify
the crucial information relevant to making predictions. Information about
the other variables, the vast majority, is totally irrelevant. Achieving
such a description requires forgetting about all microscopic details.

It is remarkable that theories that accomplish these two seemingly
contradictory goals are at all possible. They involve a very delicate
balancing act between keeping track of details, at least for a little while
(Hamiltonian evolution), and then throwing them away (projections,
coarse-graining, tracing over unwanted variables, etc.).

Our proposal cuts through this Gordian knot. If microscopic details are
truly irrelevant then the Hamiltonian evolution itself should be largely
irrelevant. The information about irrelevant details should be discarded
before, not after, it is computed. This requires formulating a dynamics
without the benefit (or, in this case, the hindrance) of Hamiltonians.

A potentially serious problem here is the loss of predictive power that
stems from the possibility of being able to choose among different dynamical
laws. What would make us prefer one law over another? Remarkably the problem
does not arise; once we define what we are talking about, namely, the states
and the criterion to distinguish among them, there is a single, unique, and
natural dynamical law that is compatible with the principle of maximum
entropy.

The views expressed here are clearly biased in favor of the information
theory approach to statistical mechanics, but they need not contradict other
points of view. The basic explanation of the second law of thermodynamics
was given by Boltzmann and Gibbs long ago but later contributions by many
authors have generated several different versions of it. The question of
which particular version is the right one remains controversial. However,
provided one adopts a certain spirit of tolerance in reading the various
authors (words such as entropy or probability can be used with very
different meanings), one sees that the different views are not always
incompatible. The point we wish to make is that irrespective of which is
one's own personal favorite reason for preferring change in the direction of
entropy increase over decrease, the \emph{same reason} should lead one to
prefer a large increase over a small one.

This applies whether we favor the information theory approach \cite{Jaynes57}%
\cite{Balian86} or one of the perhaps more traditional points of view such
as ergodic theory \cite{Lebowitz93}. For example, directing the system
toward a certain region of phase space is easier and is less sensitive to
external perturbations if the region is large than if it is small; hitting a
large target is easier than hitting a small target; that's all. Thus entropy
should increase to the maximum extent allowed by whatever constraints are
known to hold.

This last statement is widely recognized as the basis for equilibrium
thermostatics. But it shouldn't just apply to the final equilibrium state;
it should apply to every one of all the intermediate states along the
irreversible trajectory and not just to the end point. Clarifying in
precisely what sense this statement can be extended from statics to dynamics
is yet another way of stating our goals.

\section{Quantifying change}

Let the microstates of a physical system be labelled by $x$, and let $m(x)dx$
be the number of microstates in the range $dx$. We assume that a state of
the system -- that is, a macrostate -- is defined by the known expected
values $A^{{}\alpha }$ of some $n_A$ variables $a^{{}\alpha }(x)$ ($\alpha
=1,2,\ldots ,n_A$), 
\begin{equation}
\left\langle a^{{}\alpha }\right\rangle =\int dx\,p(x)a^{{}\alpha
}(x)=A^{{}\alpha }\,.  \label{Aalpha}
\end{equation}
This limited information will certainly not be sufficient to answering all
questions that one could conceivably ask about the system. Choosing the
right set of variables $\{a^{{}\alpha }\}$ is perhaps the most difficult
problem in statistical mechanics \cite{Balian}. A crucial assumption is that
Eq.(\ref{Aalpha}) is not just any random information, that it happens to be
the \emph{right} information for our purposes.

It is convenient to think of each state as a point in an $n_A$-dimensional
manifold; the numerical values $A^{{}\alpha }$ associated to each point form
a convenient set of coordinates. The principle of ME allows us to associate
a probability distribution to each point in the space of states. The
probability distribution $p(x|A)$ that best reflects the prior information
contained in $m(x)$ updated by the information $A^{{}\alpha }$ is obtained
by maximizing the entropy 
\begin{equation}
S[p]=-\int \,dx\,p(x)\log \frac{p(x)}{m(x)}.  \label{S[p]}
\end{equation}
subject to the constraints (\ref{Aalpha}). The result is 
\begin{equation}
p(x|A)=\frac 1Z\,m(x)\,e^{-\lambda _{{}\alpha }a^{{}\alpha }(x)},
\label{pzero}
\end{equation}
where the partition function $Z$ and the Lagrange multipliers $\lambda
_{{}\alpha }$ are given by 
\begin{equation}
Z(\lambda )=\int dx\,m(x)\,e^{-\lambda _{{}\alpha }a^{{}\alpha }(x)}\quad 
\text{and}\quad -\frac{\partial \log Z}{\partial \lambda _{{}\alpha }}%
=A^{{}\alpha }\,.  \label{Z and lambda}
\end{equation}
The maximized value of the entropy is 
\begin{equation}
S(A)=-\int \,dx\,p(x|A)\log \frac{p(x|A)}{m(x)}=\log Z(\lambda )+\lambda
_{{}\alpha }A^{{}\alpha }\,.  \label{S(A)}
\end{equation}

The second prerequisite to establishing that a system has changed from one
state to another is a criterion allowing us to assert that two states $A$
and $A+dA$ are not the same. Can we distinguish between the two? If $dA$ is
small enough the corresponding probability distributions $p(x|A)$ and $%
p(x|A+dA)$ overlap considerably and it is easy to confuse them. We seek a
real positive number to provide a quantitative measure of the extent to
which these two distributions can be distinguished.

The following argument is intuitively appealing. Consider the relative
difference, 
\begin{equation}
\frac{p(x|A+dA)-p(x|A)}{p(x|A)}=\frac{\partial \log p(x|A)}{\partial
A^{{}\alpha }}\,dA^{{}\alpha }.
\end{equation}
The expected value of the relative difference might seem a good candidate,
but it does not work because it vanishes identically, 
\begin{equation}
\int dx\,p(x|A)\,\frac{\partial \log p(x|A)}{\partial A^{{}\alpha }}%
\,dA^{{}\alpha }=dA^{{}\alpha }\,\frac \partial {\partial A^{{}\alpha }}\int
dx\,p(x|A)=0.
\end{equation}
However, the variance does not vanish, 
\begin{equation}
d\ell ^2=\int dx\,p(x|A)\,\frac{\partial \log p(x|A)}{\partial A^{{}\alpha }}%
\,\frac{\partial \log p(x|A)}{\partial A^{{}\beta }}\,dA^{{}\alpha
}dA^{{}\beta }\equiv g_{\alpha \beta }\,dA^{{}\alpha }dA^{{}\beta }\,\,.
\end{equation}
This is the measure of distinguishability we seek; a small value of $d\ell
^2 $ means the points $A$ and $A+dA$ are difficult to distinguish. The $%
g_{\alpha \beta }$ are recognized as elements of the Fisher information
matrix \cite{Fisher25}.

Up to now no notion of distance has been introduced on the space of states.
Normally one says that the reason it is difficult to distinguish between two
points in say, the real space we seem to inhabit, is that they happen to be
too close together. It is very tempting to invert the logic and assert that
the two points $A$ and $A+dA$ must be very close together whenever they
happen to be difficult to distinguish. Thus it is natural to interpret $%
g_{\alpha \beta }$ as a metric tensor \cite{Rao45}. It is known as the
Fisher-Rao metric, or the information metric. A disadvantage of these
heuristic arguments is that they do not make explicit a crucial property of
the Fisher-Rao metric, except for an overall multiplicative constant this
Riemannian metric is unique \cite{Amari85}\cite{Rodriguez89}.

To summarize: the very act of assigning a probability distribution $p(x|A)$
to each point $A$ in the space of states, automatically provides the space
of states with a metric structure.

The coordinates $A$ are quite arbitrary, they need not be the expected
values $\left\langle a^{{}\alpha }\right\rangle $. One can freely switch
from one set to another. It is then easy to check that $g_{\alpha \beta }$
are the components of a tensor, that the distance $d\ell ^2$ is an
invariant, a scalar. Incidentally, $d\ell ^2$ is also dimensionless. There
is, however, one special coordinate system in which the metric takes a form
that is particularly simple. These coordinates are the expected values
themselves, $A^{{}\alpha }=\left\langle a^{{}\alpha }\right\rangle $. In
these coordinates, 
\begin{equation}
g_{\alpha \beta }=-\frac{\partial ^2S(A)}{\partial A^{{}\alpha }\partial
A^{{}\beta }}\,  \label{gab}
\end{equation}
with $S(A)$ given in Eq.(\ref{S(A)}) and the covariance is not manifest.

\section{Intrinsic dynamics and time}

Our basic dynamical principle is that small changes from one state to
another are possible and do, in fact, happen. We do not explain why they
happen but, if we are given the valuable piece of information that some
change will occur, we can then venture a guess, make a prediction as to what
the most likely change will be.

Before giving mathematical expression to this principle we note that large
changes are assumed to be the cumulative result of many small changes. As
the system moves it follows a continuous trajectory in the space of states.
We almost hesitate to call this self-evident fact an assumption, but as the
example of quantum theory shows, trajectories need not exist.

Thus in order to go from one state to another the system will have to move
through intermediate states; in order to change by a distance $2d\ell $ the
system must have first changed by a distance $d\ell $.

Suppose the system was in the state $A_{old}^{{}\alpha }=A^{{}\alpha }$ and
that it changes by a small amount $d\ell $ to a nearby state. We have to
select one new state $A_{new}^{{}\alpha }=A^{{}\alpha }+dA^{{}\alpha }$ from
among those that lie on the surface of an $n_A$-dimensional sphere of radius 
$d\ell $ centered at $A^{{}\alpha }$. This is precisely what the ME
principle was designed to do \cite{Caticha00}, namely, to select a preferred
probability distribution from within a specified given set. The only
difference with more conventional applications of the ME principle is the
geometrical nature of the constraint.

We want to maximize $S(A^{{}\alpha }+dA^{{}\alpha })$ under variations of $%
dA^{{}\alpha }$ constrained by $g_{\alpha \beta }\,dA^{{}\alpha }dA^{{}\beta
}=d\ell ^2$. The notation $dA^{{}\alpha }=\dot{A}^{{}\alpha }d\ell $ is
slightly more convenient; we maximize $S(A^{{}\alpha }+\dot{A}^{{}\alpha
}d\ell )$ under variations of $\dot{A}^{{}\alpha }$ constrained by 
\begin{equation}
g_{\alpha \beta }\,\dot{A}^{{}\alpha }\dot{A}^{{}\beta }=1\,.  \label{gAA}
\end{equation}
Introducing a Lagrange multiplier $\omega $, 
\begin{equation}
\delta \left[ S(A^{{}\alpha }+\dot{A}^{{}\alpha }d\ell )-\omega \,\left(
g_{\alpha \beta }\dot{A}^{{}\alpha }\dot{A}^{{}\beta }-1\right) \right] =0,
\end{equation}
we get 
\begin{equation}
\left[ \frac{\partial S}{\partial A^{{}\alpha }}\,d\ell -2\omega \,g_{\alpha
\beta }\dot{A}^{{}\beta }\right] \delta \dot{A}^{{}\alpha }=0\,.
\end{equation}
Therefore, writing $\omega =\sigma \,d\ell /2$, we get 
\begin{equation}
\dot{A}^{{}\alpha }=\frac 1\sigma \,g^{\alpha \beta }\frac{\partial S}{%
\partial A^{{}\beta }},
\end{equation}
where $g^{\alpha \beta }$ is the inverse of $g_{\alpha \beta }$. This is our
main result; it can be rewritten as

\begin{equation}
\dot{A}^{{}\alpha }=\frac 1\sigma \,\lambda ^{{}\alpha }  \label{main1}
\end{equation}
where the vector $\lambda ^{{}\alpha }$, 
\begin{equation}
\lambda ^{{}\alpha }=g^{\alpha \beta }\,\frac{\partial S}{\partial
A^{{}\beta }}\,,
\end{equation}
is the entropy gradient. The interpretation is clear, the system moves along
the entropy gradient.

This seems such an obvious result that it can hardly be new. Notice,
however, the gradient \emph{vector} refers to the direction in which there
is a maximum increase \emph{per unit distance}; one cannot talk about the
gradient vector without having first introduced a metric. The differential
form defined by the derivatives $S_{{},\beta }=\partial S/\partial
A^{{}\beta }=\lambda _{{}\beta }$, the gradient \emph{one-form}, does not
define a direction; it is not by itself sufficient to define the trajectory.

The physical significance of the Lagrange multiplier $\sigma $ derives from
the constraint Eq.(\ref{gAA}) which, using Eq.(\ref{main1}), can be written
as 
\begin{equation}
\lambda _{{}\alpha }\lambda ^{{}\alpha }=\sigma ^2\quad \text{or}\quad
\sigma =\left( \lambda _{{}\alpha }\lambda ^{{}\alpha }\right) ^{1/2},
\end{equation}
$\sigma $ is the magnitude of the entropy gradient. Furthermore, from this
and Eq.(\ref{main1}), we get $dS=\lambda _{{}\alpha }\dot{A}^{{}\alpha
}d\ell =\sigma d\ell $,\thinspace or 
\begin{equation}
\sigma =\frac{dS}{d\ell }\,.
\end{equation}
$\sigma $ is the rate of entropy increase along the trajectory.

The main result, Eq.(\ref{main1}), determines the trajectory followed by the
system. It determines the tangent vector $\dot{A}^{{}\alpha }=dA^{{}\alpha
}/d\ell $, but not the ``velocity'' $dA^{{}\alpha }/dt$. To fix this
something must be said about the universe external to the system, something
that relates the distance $\ell $ relative to the external time $t$. This
is, in part, the role normally played by the Hamiltonian, it fixes the
evolution of a system relative to external clocks. If we cannot appeal to
such information (presumably because we do not have it, but perhaps because
we just do not want to), then the only ``time'' available must be internal
to the system, intrinsic to the geometry of the space of states.

One convenient choice of intrinsic time $\tau $ is the distance $\ell $
itself, or $d\tau =d\ell $. Intrinsic time is change. The equation of motion
is very simple: the trajectory, $A^{{}\alpha }=A^{{}\alpha }(\tau )$, is
along the entropy gradient, and the system moves with unit velocity, $\dot{A}%
^{{}\alpha }\dot{A}_{{}\alpha }=1$, or $d\ell /d\tau =1$.

The absolute speed $d\ell /dt$ remains unknown. Interestingly, there is no
guarantee that $\tau $ will elapse relative to our own external $t$, we
could have a situation with $d\tau /dt=0$. A pile of sand could, if left
alone, just stay at $A^{{}\alpha }(\tau _0)$ forever; its intrinsic time $%
\tau $ has stopped at $\tau _0$. The pile does not change, because it did
not have (intrinsic) time to change. (One can play endless word games here.)

However, should a measurement of one of the variables, for example $A^{{}1}$%
, indicate a change from the value $A^{{}1}(\tau _0)$ to the value that one
would normally associate with another state along the trajectory, say the
value $A^{{}1}(\tau _1)$ at the later time $\tau _1$, then one is
immediately led to infer that the system has moved along the trajectory.
Most probably all the other variables have also changed from $A^{{}\alpha
}(\tau _0)$ to $A^{{}\alpha }(\tau _1)$. In this case the variable $%
A^{{}1}(\tau )$ is playing the role of an internal clock. The variable $%
A^{{}1}$ is a good clock provided one can invert $A^{{}1}=A^{{}1}(\tau )$,
to get $\tau =\tau (A^{{}1})$. Then, the changes in all other variables $%
A^{{}\alpha }=A^{{}{}\alpha }[\tau (A^{{}1})]=A^{{}{}\alpha }(A^{{}1})$ can
be referred, correlated to the change in $A^{{}1}$. We see that the loss of
predictive power due to the unknown absolute speed $d\ell /dt$ is quite
minimal, particularly for high dimensionality (large $n_A$).

At this point one could agree that the notion of $\tau $ is useful, perhaps
even elegant. But are we justified in \emph{calling} it time? Perhaps these
are mere word games, but if we do call $\tau $ time, then being a distance
it provides us with a model of duration. Furthermore, the very definite
ordering of states along the trajectory $A^{{}\alpha }(\tau )$ provides a
realization of a temporal order. Finally, the dynamics is intrinsically
asymmetric; the trajectory is intrinsically oriented. There is one direction
in which entropy increases providing a clear distinction between earlier and
later. So this is our answer: we are justified in calling $\tau $ time,
because if we do, then we have a neat model, an explanation for temporal
order, for time asymmetry, and for duration. What better reasons do we need?

We close this section with the observation that the system does not follow a
geodesic in the space of states. From Eq.(\ref{main1}) we can show that the
acceleration vector, given by the absolute derivative (we assume a
Riemannian geometry, with the Levi-Civita connection) 
\begin{equation}
\frac{D\dot{A}^{{}\alpha }}{d\tau }=\dot{A}_{;\beta }^{{}\alpha }\,\dot{A}%
^{{}\beta }=g^{\alpha \beta }\,f_{\beta \gamma }\,\dot{A}^{{}\gamma },
\end{equation}
does not vanish. The ``thermodynamic force'' resembles the Lorentz force law
in electrodynamics. The ``field strength'' tensor $f_{\alpha \beta }$, given
by $f_{\alpha \beta }=\dot{A}_{{}\alpha ;\beta }-\dot{A}_{\beta ;\alpha }$,
is antisymmetric as needed to preserve the unit magnitude of the velocity $%
\dot{A}^{{}\alpha }$.

\section{Reciprocal relations}

The standard theory of irreversible thermodynamics, due to Onsager \cite
{Onsager31}, is based on the usual postulates of equilibrium thermostatics
supplemented by the additional postulate that the microscopic laws of motion
are symmetric under time reversal. A brief ouline is the following.

As the system moves along its trajectory entropy increases at a rate 
\begin{equation}
\frac{dS}{dt}=\frac{\partial S}{\partial A^{{}\alpha }}\frac{dA^{{}\alpha }}{%
dt}=\lambda _{{}\alpha }\frac{dA^{{}\alpha }}{dt}
\end{equation}
relative to the external time $t$; the variables $\lambda _{{}\alpha }$ are
called thermodynamic forces, and $dA^{{}\alpha }/dt$ are called fluxes. In
this theory linear relations between fluxes and forces are postulated,

\begin{equation}
\frac{dA^{{}\alpha }}{dt}=L^{\alpha \beta }\lambda _{{}\beta }\,,
\end{equation}
for which there is abundant experimental evidence, at least close to
thermodynamic equilibrium.

The significance of these relations lies in that they postulate crossed
connections between a flux of type $\alpha $ and a force of type $\beta $,
and vice versa. (Thus, a temperature gradient will not just generate an heat
current; it may also generate electric currents, matter flows, and so on.)
The strength of these effects is measured by the phenomenological Onsager
coefficients $L^{\alpha \beta }$. The central result of the theory is the
reciprocal relation between these crossed effects. The reciprocity theorem,
proved by Onsager on the basis of microscopic reversibility, states that the
matrix of phenomenological coefficients is symmetric 
\begin{equation}
L^{\alpha \beta }=L^{\beta \alpha }\,.
\end{equation}

The intrinsic dynamics discussed in the previous section also leads to
reciprocal relations. The equation of motion, Eq.(\ref{main1}), gives 
\begin{equation}
\frac{dA^{{}\alpha }}{dt}=\frac{d\tau }{dt}\frac{dA^{{}\alpha }}{d\tau }=%
\frac{d\tau }{dt}\frac 1\sigma \,g^{\alpha \beta }\lambda _{{}\beta }\,.
\label{dA/dt}
\end{equation}
This allows us to identify the Onsager coefficients as 
\begin{equation}
L^{\alpha \beta }=\frac{d\tau }{dt}\frac 1\sigma \,g^{\alpha \beta }.
\end{equation}
These coefficients are not constants, they vary along the trajectory, $%
L^{\alpha \beta }=L^{\alpha \beta }(A)$.

What is interesting here is that their symmetry follows from the symmetry of
the metric tensor. No hypothesis about microscopic reversibility was needed;
in fact, microscopic dynamics was not mentioned at all. In addition, the
validity of Eq.(\ref{dA/dt}) is not restricted to the immediate vicinity of
equilibrium. To the extent that the variables $A$ are the right variables to
describe phenomena far from equilibrium, the reciprocal relations should
still hold.

\section{Dynamics constrained by conservation}

Beyond the fact that changes happen, perhaps the most common additional
information that one can have about an irreversible process is that some
quantities are conserved. As an illustrative example we consider two systems
that are allowed to exchange some conserved quantities and evolve towards
equilibrium. To fix ideas we could think of an ideal gas filling two vessels
at different temperatures and chemical potentials. Once the two vessels are
connected, for example by a tube, a little hole, or a a porous plug, matter
and energy will flow until equilibrium is reached.

To keep this as simple as possible we assume the experimental conditions are
such that throughout the process the two systems remain homogeneous and
independent. The first system is described by variables $A^{{}\alpha }$, the
second is described by primed variables $A^{\prime {}\alpha }$, and the
entropies, given by Eq.(\ref{S(A)}), are additive 
\begin{equation}
S_T(A,A^{\prime {}})=S(A)+S^{\prime }(A^{\prime {}}).
\end{equation}
Since the quantities $A$ are conserved the dynamics is constrained by $%
A^{\prime }=A_T-A$, with $A_T$ fixed, or $\dot{A}^{\prime }=-\dot{A}$. The
conservation constraint could be incorporated using Lagrange multipliers;
for this simple example it is just as easy to eliminate $A^{\prime }$.

In our ideal gas example, the variables could be energy, $A^1=E$, and number
of molecules, $A^2=N$. This crucial part in setting the problem, choosing
the description, is the one most likely to go wrong. If the hole coupling
the two vessels is too large, the ME predictions below will fail. The
failure is not to be blamed in the ME method, but on the choice of
variables: the pair $E$, $N$ is not enough to codify the relevant
information of say, a turbulent flow. The same remark applies if the
connecting porous plug is such that heat can be easily exchanged but there
is resistance to matter flow. In this case additional variables are needed,
perhaps describing the physical state of the plug and the gas in it.

Suppose the system was in the state $A$ and that it changes by a small
amount $d\tau $ to a nearby state. To select one new state $A+\dot{A}d\tau $
from among those that lie on the surface of a sphere of radius $d\tau $
centered at $A$, we maximize 
\begin{equation}
S_T(A+\dot{A}d\tau )=S(A+\dot{A}d\tau )+S^{\prime }(A_T-A-\dot{A}d\tau )
\end{equation}
under variations of $\dot{A}^{{}\alpha }$ constrained by 
\begin{equation}
g_{\alpha \beta }\,\dot{A}^{{}\alpha }\dot{A}^{{}\beta }=1\,.
\end{equation}
where the Fisher-Rao metric, Eq.(\ref{gab}), is given by 
\begin{equation}
g_{\alpha \beta }=-\frac{\partial ^2}{\partial A^{{}\alpha }\partial
A^{{}\beta }}\left( S(A)+S^{\prime }(A_T-A)\right) .
\end{equation}
The result is 
\begin{equation}
\frac{dA^{{}\alpha }}{d\tau }=\frac 1\sigma \,g^{\alpha \beta }\left(
\lambda _{{}\beta }-\lambda _{{}\beta }^{\prime }\right) ,
\end{equation}
where $\sigma $ is the rate of entropy production $\sigma =dS_T/d\tau $. The
system evolves until the conjugate variables $\lambda $ are equalized.

\section{Final remarks}

The main conclusion is simple: unless there is positive evidence to the
contrary, our best prediction is that the system evolves along the entropy
gradient. What is perhaps not so trivial is that, unlike other conventional
forms of dynamics, this intrinsic dynamics does not require an additional
postulate. It is the unique dynamics that follows from the maximum entropy
principle and nothing else. Another nontrivial aspect is that the model
supplies its own notion of time. Since the irreversible macroscopic motion
is not explained in terms of a reversible microscopic motion there is no
need to explain irreversibility, this question never arises. Similarly,
there is no need to explain the second law of thermodynamics; it is the
second law (in the form of the ME axioms) that explains everything else.

These ideas can be explored further in a number of directions. There is, for
example, the relation with other theories of irreversible processes, such as
the equations of hydrodynamics. Another possibility is to extend the theory
to account for fluctuations and diffusion. The intrinsic dynamics proposed
above is deterministic, but to the extent that the ME principle does not
completely rule out distributions of lower entropy \cite{Caticha00},
fluctuations about equilibrium and about the deterministic motion are
possible.

Perhaps the most intriguing question to pursue stems from the possibility of
deriving dynamics from purely entropic arguments. This is clearly valuable
in areas where the microscopic dynamics may be too far removed from the
phenomena of interest, say in biology or ecology, or where it may just be
unknown or perhaps even inexistent, as in economics. One could argue that
these theories would be phenomenological as opposed to fundamental, that
within physics the search for a fundamental mechanics would still be left
open. However, in previous work we have shown \cite{Caticha99} that entropic
arguments do account for a substantial part of the formalism of quantum
mechanics, a theory that is presumably fundamental. Perhaps the fundamental
theories of physics are not so fundamental; they are just consistent,
objective ways to manipulate information.

\end{document}